\documentclass[twocolumn]{revtex4} 
\usepackage{graphicx}

\newcommand{\be}{\begin{equation}}
\newcommand{\ee}{\end{equation}}

\newcommand{\tg}{T_{\rm g}}
\newcommand{\tk}{T_{\rm K}}
\newcommand{\Tx}{T_{\rm x}}
\newcommand{\dcv}{\Delta c_v}

\begin{document}

\title{Droplets and the Configurational Entropy Crisis for Random First Order Transitions}

\author{M. P. Eastwood}
\author{P. G. Wolynes}
\affiliation{Department of Chemistry and Biochemistry, University of California, 9500 Gilman Drive,
La Jolla, CA 92093-0371, USA}

\date{\today}

\begin{abstract}
We consider the effect of droplet excitations in the random first order transition theory
of glasses on the configurational entropy. The contribution of these excitations is estimated both at and above the ideal
glass transition temperature. The temperature range where such excitations could conceivably
modify or ``round-out'' an underlying glass transition temperature is estimated, and found
to depend strongly on the surface tension between locally metastable phases in the supercooled
liquid. For real structural glasses this temperature range is found to be very narrow, consistent with the
quantitative success of the theory. For certain finite-range spin-glass
models, however, the surface tension is estimated to be significantly lower leading to much stronger
entropy renormalizations, thus providing an explanation for the lack of a strict thermodynamic glass transition in
simulations of these models.
\end{abstract}

\maketitle

A number of models of disordered glassy systems such as random heteropolymers \cite{shakhnovich89}, 
$p$-spin glasses \cite{gardner85} and Potts
spin glasses \cite{gross85} in the mean field limit undergo a thermodynamic transition at which an appropriately
defined configurational entropy vanishes. While supercooled liquids in the laboratory never reach such
a transition before falling out of equilibrium, rather natural extrapolations of the experimentally
measured thermodynamics also suggest that if a fluid could remain equilibrated upon slow cooling, an 
entropy crisis would occur at an ideal glass transition temperature $\tk$ \cite{kauzmann48}. This temperature is remarkably
near the temperature where, again a natural extrapolation of the experimental kinetic data suggests that
relaxations become infinitely slow.

Over many decades, contemplating such an ideal glass transition has entered into discussions
of glassy dynamics. The ``random first order transition" (RFOT) exhibited by the mean-field models has become
a central element of the modern theory of glasses 
\cite{kirkpatrick87,kirkpatrick87b,kirkpatrick87c,kirkpatrick89,mezard99,mezard00}. Recently this theory has been shown to 
predict quantitatively from a microscopic starting point the dynamics of supercooled liquids in the laboratory
\cite{xia00,xia01b},
including the relaxation time distribution and the length scale of dynamical heterogeneity \cite{xia01}. 
This theory
also leads to a natural way to describe the anomalous thermodynamic and transport properties of low
temperature glasses, when the theory is quantized \cite{lubchenko01}. Despite its venerable history 
\cite{adam65} the idea of an underlying
ideal glass transition related to an entropy crisis has also been greeted by much controversy since it requires
extrapolation from extant data. 
Yet only one \textit{a priori} objection to an entropy crisis has been crisply formulated: configurational entropy cannot
vanish above absolute zero since a system with finite range forces can always support point defects at
any finite temperature \cite{stillinger88}. The most naive incarnation of this objection, the existence of 
vacancies or interstitials, can easily be dismissed by practical persons. At the densities of glasses the
concentration of such defects is miniscule. Reasonable estimates suggest they occur in 1 part in $10^4$ so they
hardly contribute to the configurational entropy, which is typically $1k_B$ per particle at the 
laboratory liquid-glass transition temperature, $\tg$. More troubling is that the random first order transition theory itself
requires the existence of excitations (``entropic droplets") which are localized and have finite energy cost.
It is conceivable these excitations can smear out or modify the critical exponents of the transition predicted
by mean-field theory. In this paper, we estimate what the droplet contribution to the entropy both below and
above an ideal glass transition would be. This calculation parallels Fisher's droplet calculations on the   
essential singularities in the thermodynamics encountered near an ordinary first order phase transition \cite{fisher67}. The
calculation shows how close to an ideal glass transition we must go to see these renormalization effects and
quantifies where the notion of configurational entropy becomes ambiguous. In essence this calculation provides 
something like a
``Ginzburg" criterion \cite{ginzburg60} describing the breakdown of mean-field
theory for a random first order transition. The size of the strongly renormalized regime depends strongly
on the effective surface tension between mean-field replica symmetry broken phases. The effective surface tension
parameter is actually quite large for the liquid-glass transition because it depends on the Lindemann vibrational amplitude which is so small.
Effectively there is, then, a small parameter in the theory which allows 
the mean-field estimates to prevail until very close to the transition
(much as does the BCS theory of conventional superconductivity). The entropy crisis for liquids will
be noticeably avoided only when the residual entropy is of the order $10^{-4}k_B$ and $(T-T_K)/T_K$ is approximately $10^{-4}k_B/\dcv$ where
$\dcv$ is the mean-field heat capacity discontinuity at the transition. In contrast, the effective surface tension
for a 10-state nearest neighbour lattice Potts glass is quite small. This small value is consistent with the avoidance of an
entropy crisis in a recent simulation of this system \cite{brangian02}. Since the surface tension parameter
also tunes the degree of slowing as the transition is approached understanding these effects may allow
the design of simulations that can more effectively confront the theory.

We begin by very briefly summarizing the essential points of the RFOT theory of glasses, and
then
construct a simple partition function for the droplet excitations.
After discussion of droplet effects at $\tk$, we evaluate the temperature range over 
which such excitations will be significant for a structural glass as $T$ approaches
$\tk$ from above. Finally, we apply the results to several glassy systems
with different surface tensions.

Below a temperature $T_A$, which corresponds to the dynamical glass
transition predicted by mode-coupling theory \cite{gotze91}, reconfiguration of a real
supercooled liquid occurs by activated processes.
According to the RFOT theory of glasses
for $\tk<T<T_A$ the driving force for reconfiguration of a locally frozen region is the configurational
entropy, while there is an energetic cost due to the surface tension between two different locally
metastable solutions. The free energy profile for reconfiguration of a region of radius $r$ is thus
\be
\label{eq:F}
F(r)=4\pi r^2\sigma(r)-\frac{4}{3}\pi r^3\rho TS_c
\ee
where $S_c$ is the configurational entropy density per molecule, and $\rho$ is the number density.
 For the
experimental comparison the measured $S_c$ was used, but the theory more formally would call for its mean-field value.
Generalization of a renormalization group (RG) argument due to Villain \cite{villain85} suggests that the surface tension
is reduced from its bare value $\sigma_0$ (obtained between two typical metastable states) due to wetting
of the interface by additional metastable states, according to $\sigma(r)=\sigma_0(r_0/r)^\frac{1}{2}$, with 
$r_0=\rho^{-\frac{1}{3}}$ a typical intermolecular distance \cite{kirkpatrick89}. 
It is helpful to express the two key parameters (surface tension and configurational entropy)
in terms of dimensionless parameters that are approximately constant as $T\rightarrow\tk$. Specifically,
 the configurational entropy is assumed to vanish linearly, $S_c=\frac{T-\tk}{\tk}\frac{\dcv}{k_B}$, where
$\dcv$ is the heat capacity jump per particle at the transition, and we define a reduced surface tension
$A=4\pi r_0^2\sigma_0(\tk)/k_B\tk$.
 In the region $\tk<T<T_A$, we assume that $4\pi r_0^2\sigma_0(T)/k_BT$ is slowly changing, and
may be approximated by $A$. The reconfiguration barrier may then be written in Vogel-Fulcher form
as $\beta\Delta F^\ddagger=D\tk/(T-\tk)$, with the fragility $D=\frac{3}{16\pi}\frac{A^2}{\dcv/k_B}$.
The supercooled liquid is viewed as a mosaic of metastable regions of typical size
$R/r_0=\left(\frac{3A}{4\pi}\frac{k_B}{\dcv}\right)^\frac{2}{3}\left(\frac{\tk}{T-\tk}\right)^\frac{2}{3}$,
as follows from $F(R)=F(0)$.
For structural glasses $\dcv/k_B$ varies a great deal from substance to substance over a typical range of 0.2 to 5.
However, a density functional calculation \cite{xia00}
gives $A\approx23$ for structural glasses approximately independently of molecular details since it 
depends only logarithmically on the Lindemann ratio of mean-square vibrational amplitude to 
near-neighbour spacing which characterizes the vibrations at $\tg$. 

\begin{figure}[th!]
\includegraphics[width=7.0cm]{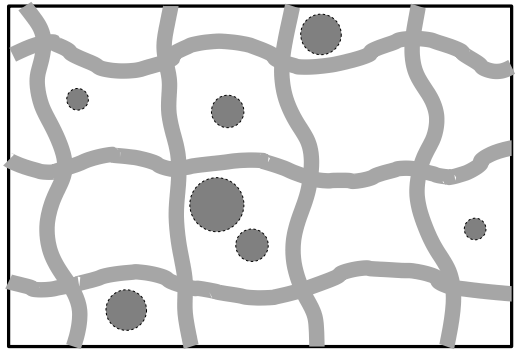}
\caption{\label{fig:mosaic} A schematic of the structure of a supercooled liquid near $\tk$, showing
droplet excitations within the mosaic cells.}
\end{figure}

Within each domain of the mosaic there should exist further droplet excitations with free energy
profiles also given by eq.\ (\ref{eq:F}), as shown schematically in fig.\ \ref{fig:mosaic}. 
Indeed, since these droplets may also contain even smaller
droplets, this suggests that like the surface tension the configurational entropy should be determined by RG equations.
In this paper, we estimate the region where such a renormalization
should be significant and leave the full RG calculation for future work, although we note 
that in the case of the random field Ising model, the presence of droplets within droplets is not
sufficient to raise the lower critical dimension from $d=2$ to $d=3$ \cite{imbrie84,imbrie85}.
The partition function for droplet formation within a single mosaic cell is
$Z=\sum_{\{n_j\}}\exp(S_{comb}(\{n_j\})-\beta\sum_in_iF(r_i))$,
where the free energy
cost of forming a droplet of radius $r_i$ about a given position is given by eq.\ (\ref{eq:F}), and $S_{comb}$ is an entropy associated
with the different positions the droplets may take up within the domain. 
Assuming that there is approximately one distinct droplet position for each molecule 
that a droplet may be centered on, gives $\frac{1}{n_i!}(\rho V_{eff})^{n_i}$ 
states of $n_i$ identical independent droplets. The effective volume within a domain 
in which a droplet may move, $V_{eff}=\frac{4}{3}\pi R_{eff}^3$, is taken to be smaller than the domain 
volume due to the presence of the domain walls of finite thickness. While strictly speaking the RG wetting
argument suggests $R_{eff}\propto R$ \cite{kirkpatrick89}, with essentially no effect on the results we 
take $R_{eff}=R-r_0$.
Excluded volume effects
have been entirely ignored, even between identical droplets, as has the reduction in the volume
available to position larger droplets due to the restriction that the entire droplet
must lie within the domain walls (this latter effect may be straightforwardly included, but 
is not significant for parameters where ignoring the excluded volume is a good approximation).
Under these approximations the entropy $S_{comb}$ reduces to $S_{comb}=\sum_i\left\{n_i\ln[\rho V_{eff}]-\ln n_i!\right\}$,
and the partition function may easily be evaluated giving a free energy density of independent droplets 
$F_{drop}=-(k_BT/\rho V)\ln Z=-kT(V_{eff}/V)\sum_i e^{-\beta F_i}$. 
Droplets may be distinguished by the number of molecules 
they contain, so what is meant by the sum over droplet types $i$ in eq.\ (\ref{eq:Sdrop}) is
$i=1,2,\dots,i_{max}$, where $i_{max}$ is the maximum size of a droplet that can
fit inside the domain $i_{max}\le \rho V_{eff}$; alternatively ---and no more approximately--- we may make the
replacement $\sum_{i=1}^{i_{max}}\rightarrow4\pi\rho\int_{r_s}^{R_{eff}}r^2dr$, with $r_s=\left(\frac{3}{4\pi}\right)^{\frac{1}{3}}r_0\approx0.62r_0$.
The corresponding droplet entropy density (which follows from differentiation with $S_c$ and $\sigma_0$ fixed) is
\be
\label{eq:Sdrop}
S_{drop}/k_B=4\pi\rho\frac{V_{eff}}{V}\int r^2(1+A(r/r_0)^\frac{3}{2})e^{-\beta F(r)}dr.
\ee

At $\tk$ the droplet entropy density
$S_{drop}(\tk)/k_B=(8\pi/3)\left[\tilde{r}_s^3+3\tilde{r}_s^{\frac{3}{2}}/A+3/A^2\right]
\exp(-A\tilde{r}_s^{\frac{3}{2}})$,
where $\tilde{r}_s=r_s/r_0$. For a structural glass, with the values given above, $S_{drop}(\tk)/k_B\approx
3\times10^{-5}$. For structural glasses, the contribution to the droplet entropy comes overwhelmingly from
the smallest droplets and is thus a defect-like entropy. Unfortunately, on the short length scales of such 
droplets (perhaps
involving only a single molecule) the calculation of the energetic cost in terms of a surface
tension becomes less accurate and the details of the particular molecules may become relatively
more important. The effect of the uncertainty in the energetic cost of forming small droplets on
the droplet entropy may be allowed for by permitting $\tilde{r}_s$ to vary in the range 0.4 to 1.0, 
which corresponds to the energy of the smallest droplets being a factor of two smaller or larger than 
estimated using $\tilde{r}_s=0.62$.
Even allowing for this uncertainty the resultant droplet entropy at $\tk$ remains tiny, in the range $1\times10^{-9}$ to 
$3\times10^{-3}k_B$.

\begin{figure}[th!]
\includegraphics[width=7.0cm]{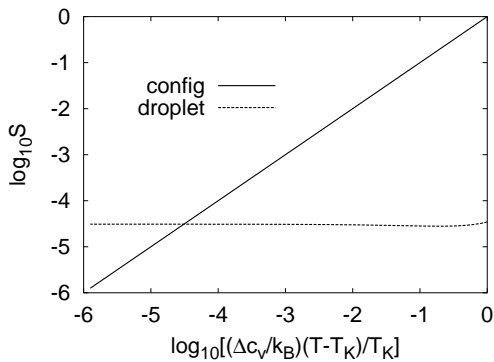}
\caption{\label{fig:S} The configurational entropy and the droplet correction to the
entropy for a structural glass as a function of temperature in the temperature range $(T-\tk)/\tk=10^{-6}k_B/\dcv$
to $k_B/\dcv$.}
\end{figure}

\begin{figure}[th!]
\includegraphics[width=7.0cm]{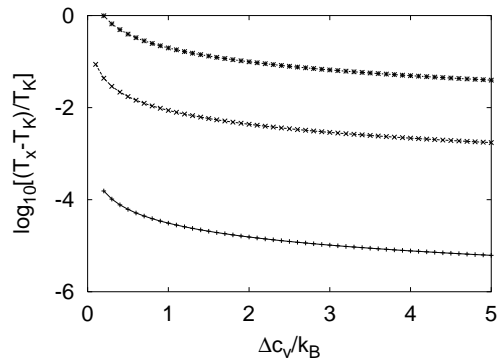}
\caption{\label{fig:Tcross} $\log_{10}[(\Tx-\tk)/\tk]$, where $\Tx$ is the crossover temperature
 as a function of the heat
capacity jump, for $A=23$ (bottom; appropriate for a structural glass), $A=12$ (middle; as may
be appropriate for a protein-like random heteropolymer), and $A=6$
(top; possibly appropriate for some spin models as discussed in the text).}
\end{figure}

We now consider $T>\tk$ and estimate the crossover temperature, $\Tx$, where the droplet entropy
and configurational entropy become comparable. 
For $\tk<T<\Tx$ 
droplet effects may modify the character of the random first order transition. 
As $S_{drop}(\tk)$ is finite and small, it is clear that $\Tx>\tk$
exists and since, as fig.\ \ref{fig:S} shows, $S_{drop}(T)$ is
not rapidly varying as $T$ approaches $\tk$, $\Tx$ may be estimated from from
$S_{drop}(\tk)=S_c(T)$. For large values of the reduced surface tension and $\tilde{r}_s=0.62$ this
leads to 
$\log_{10}\left[(\dcv/k_B)\times(\Tx-\tk)/\tk\right]=-0.21A+0.3$.
For structural glasses this suggests
$(\Tx-\tk)/\tk$ lies in the range $1.5\times10^{-4}$
(for strong liquids) to $6\times10^{-6}$ (for fragile liquids). Such cross-over temperatures lie well below those
accessible to equilibrium measurements, since the
$\alpha$ relaxation times at $\Tx$ will be of the order $\tau_0\exp(10^6)$, effectively infinite. 
We show in fig.\ \ref{fig:Tcross} the cross-over temperature as a function
of $\dcv$ over a range of $\dcv$ appropriate to a structural glass. 
The widening of the renormalized region with decreasing $\dcv$ (increasing fragility) is clear.
The conclusion that $\Tx$ is inaccessible to direct experimental study for structural glasses appears
robust to the uncertainties in the energy cost calculation: even taking $\tilde{r}_s=0.38$
(corresponding to an energy cost of half the natural estimate) leads to $(\dcv/k_B)\times(\Tx-\tk)/\tk
=2\times10^{-3}$ while $(\dcv/k_B)\times(\tg-\tk)/\tk\approx1$.

\begin{figure}[th!]
\includegraphics[width=7.0cm]{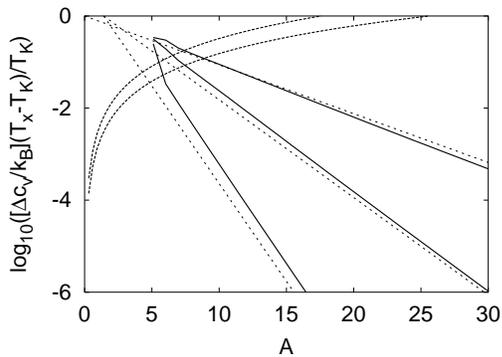}
\caption{\label{fig:lambda} Logarithm of the width of the renormalized region
as a function of the dimensionless
surface tension parameter.
The three solid curves correspond (from top to bottom) to $\tilde{r}_s=0.38$, 0.62 and 0.98,
with $\Tx$ calculated from $S_c(\Tx)=S_{drop}(\Tx)$ and eq.\ (\ref{eq:Sdrop}).
The three dotted lines are the asymptotes $\ln\left[(\dcv/k_B)\times(\Tx-\tk)/\tk\right]=-\tilde{r}_s^\frac{3}{2}A
+\ln\left[8\pi\tilde{r}_s^3/3\right]$. The two dashed lines
give the laboratory glass transition temperature as $\log_{10}[(\dcv/k_B)\times(\tg-\tk)/\tk]$
for waiting times appropriate to experiments
on structural glasses (lower) and computer simulations on spin-glasses (upper).}
\end{figure}

While structural glasses all have $A\approx23$ other glassy systems are expected
to have different bare surface tensions.  The width of the renormalized region depends
strongly on $A$.  
Figure \ref{fig:lambda}
gives the full dependence of the width of the renormalized region on the bare surface tension 
parameter $A$.
Also shown is the position of the laboratory glass transition temperature $T_g$, defined
in terms of the experimental and microscopic times via
$\tau_g/\tau_0=\exp(D\tk/(T_g-\tk))$ with both $\tau_g/\tau_0=10^{17}$ (applicable
to experiments on structural glasses with relaxations in the hours range) and $\tau_g/\tau_0=10^{8}$ (typical
of a computer simulation of a spin glass). For large values of the surface tension ($A>15$) $\Tx$
is found to lie significantly below $T_g$ and the additional droplet entropy is small at
any temperature where equilibration can be achieved in practice. For surface tensions in the
range $5<A<10$, on the other hand, the values of $\Tx$ and $T_g$ will be more comparable and
the effects of droplet excitations should become apparent. Depending on the details of the
model and the timescale accessible to the experimenter as $A$ is decreased from 10 to 5, at some point
$\Tx$ and $T_g$ will cross and the effect of the droplet renormalization may be large. We note
that by $A=6$ the independent droplet approximation breaks down since $\sim30\%$ of the 
volume of domains is predicted to be filled with droplet excitations at $\Tx$. Nevertheless,
this does not change the conclusion that droplets are significant at this surface tension.

To use the droplet entropy renormalization criterion we need to know
$A$ for systems with finite-ranged interactions that undergo
RFOTs in the mean-field limit. 
Unfortunately, for protein, Potts and $p$-spin models accurate calculations of $A$ are somewhat involved.
Therefore, here we content ourselves to estimate the value of $A$ for various models
using  rough calculations, that should be sufficiently 
accurate to determine whether or not droplet excitations are likely to play an important
role in the entropy. We first make an estimate for polymeric, and then some spin-glass models.

The surface tension for a random heteropolymer may be estimated using the argument of Takada
and Wolynes \cite{takada97}, or alternatively by approximating
the surface tension between two neighboring misfolded regions of a random heteropolymer with the
surface tension of a sharp interface between folded and unfolded regions of a protein. The latter 
quantity is a key quantity in the capillarity model of protein folding \cite{wolynes97,finkelstein97},
which in its simplest form assumes folding occurs by nucleation with a sharp interface
between completely folded and completely unfolded regions. At the folding temperature,
$T_f$, the capillarity model predicts a free energy barrier of $\Delta F/k_BT_f=\frac{4}{27}
\left(\frac{3}{4\pi}\right)^{\frac{2}{3}}AN^\frac{2}{3}$. A 
knowledge of the variation of folding rates at $T_f$ with protein size will thus give an estimate
of the reduced surface tension. Such a study has been performed by Koga and Takada \cite{koga01},
who find for simulations of model proteins a folding rate that varies as $\log_{10}k_f\approx-0.5cN^\frac{2}{3}$,
with $c$ a roughly $N$-independent constant that depends on protein topology called relative contact order. 
$c$ measures how frequently the chain returns to itself.
The same relation holds for real proteins but folding times vary over more than four more orders of magnitude
and the coefficient is large,
$\log_{10}k_f\approx-2.3cN^\frac{2}{3}$. $c$ is in the range 0.08 to 0.2 
giving a range of values for $A$ from 7 to 19 for a real proteins in the laboratory,
 suggesting that droplet renormalization is
not significant for the majority of cases. For the simulation models, on the other
hand, the calculation gives $A$ in the range 1.6 to 4 indicating that droplet renormalization is likely
to be very significant for off-lattice models of random heteropolymers.

\begin{table}[th!b]
\caption{Estimates of the reduced surface tension parameter $A$ for some models
studied by simulation. $A_{sp}$ is the sharp interface approximation, while $A_{\xi}$
and $A_\tau$ are the results of extracting $A$ from the simulation data (using the RFOT
theory) on the correlation length and relaxation times respectively.}
\begin{tabular}{ccccc}
model & $k$ & $A_{sp}$ & $A_{\xi}$ & $A_\tau$ \\
\hline
     $p=4$ $M=3$ \cite{campellone99,franz99} & 1/6 & 10 & 4.8 & -- \\
     $p=4$ plaquette \cite{alvarez96}& 1/3 & 10 & -- & 3.3 \\
     $q=10$ Potts \cite{brangian02}& 1/6 & 8 & -- & -- \\
\end{tabular}
\end{table}

Computer simulations have been performed on a few spin-glass systems 
that exhibit a one-step RSB transition in the mean-field limit including  $p=4$ spin
models on different lattices \cite{franz99,alvarez96,rieger92} 
and $q=10$ state Potts glass \cite{brangian02}. Unlike 
the $p=4$ spin models the Potts model did not display the characteristic signs of a glass transition. An
estimate for $A$ for these systems based on the sharp short-range interface argument is clouded
by the nearness of the mean-field $T_A$ and $\tk$ for these models. 
If we nevertheless assume a sharp planar interface, and take the free energy cost of forming the interface
to be approximately one half of the energetic cost of randomizing the interactions across the interface, 
we would find $r_0^2\sigma_0=k\left\vert E_k\right\vert$ (or $A=4\pi k\left\vert E_{\rm K}\right\vert/\tk$) 
with $k$ the fraction of interactions per site that are broken in forming the interface and $E_{\rm K}$ the
energy per site at $\tk$.
The resulting estimates for $A$ using $E_{\rm K}$ and $\tk$ from simulations (or from mean-field
theory in the case of the Potts model)  are shown
in the table. Taken at face value, these suggest that the $p=4$ spin models have a surface tension
only just sufficient to avoid substantial corrections to mean-field theory, while there may be significant
renormalization for the 10-state Potts model.
Alternatively an estimate
of the surface tension
can be obtained using the observed growth in correlation length \cite{franz99} and relaxation
times \cite{alvarez96} which indeed have the functional forms predicted by RFOT theory.
In doing this it is necessary to assume the theoretical results
apply outside the region $\tk< T\ll T_A$.
These estimates, also given in the table are much lower.
These low values are consistent with the observation that
$T_A$ (where
surface tension vanishes in finite ranged models) lies only slightly above $\tk$ for these models in the
mean-field limit; in particular $T_A/\tk=1.009$
for the 10-state Potts glass \cite{desantis95}. We conclude that the apparent lack of a glass
transition in the 10-state Potts model is due to a low effective microscale surface tension. 
Since $\left\vert E_{\rm K}\right\vert/\tk=2S_0$ (in a random energy model, 
with $S_0$ the total entropy per site) it can be anticipated that $A$ will
increase approximately logarithmically with the number of states in the Potts model but proportional to $p$
in a $p$-spin model  \cite{desantis95}.
Achieving spin systems with a surface
tension similar to that in a structural glass would require a $p\approx20$ ($M=p-1$) spin model, but a $q=10^3$ 
(or even higher) state Potts glass. Unfortunately spin models that are relatively easily
simulated appear
to have substantial corrections to the current form of RFOT theory while real structural glasses and protein-like
random heteropolymers have much smaller corrections.
\acknowledgments
This work was supported by NIH grant 5 R01 GM44557.

\bibliography{glass}

\end{document}